
%

\input harvmac
\input tables
\tolerance=5000
\def\bigfirst#1#2{
\ifx\answ\bigans#1%
\else#2%
\fi}

\def\np#1#2#3{Nucl. Phys. {\bf #1} (#2) #3}
\def\pl#1#2#3{Phys. Lett. {\bf #1} (#2) #3}
\def\prl#1#2#3{Phys. Rev. Lett. {\bf #1} (#2) #3}
\def\pr#1#2#3{Phys. Rev. {\bf #1} (#2) #3}
\def\prd#1#2#3{Phys. Rev. {\bf D#1} (#2) #3}

\def\nuvc#1#2#3{Nuovo Cimento {\bf #1A} (#2) #3}
\def\blankref#1#2#3{   {\bf #1} (#2) #3}
\def\ibid#1#2#3{{\it ibid,\/}  {\bf #1} (#2) #3}
\def\ie{{\it ie}}

\def\ccdot{\hbox{\kern-.1em$\cdot$\kern-.1em}}
\def\a{\alpha}
\def\ae{\alpha_{\rm e.m.}}
\def\as{\alpha_s}
\def\ag{\alpha_{GUT}}

\def\ct{\cos^2\theta}
\def\st{\sin^2\theta}
\def\MM{\tilde M}
\def\tb{\tan\beta}

\def\dl{\delta_{\rm light}}
\def\gf{\gamma^5}

\def\ma{m_{\rm avg}}
\def\MG{M_{\rm GUT}}

\def\MS{M_{\rm SUSY}}
\def\dg{\delta g}
\def\dv{{\delta v}}
\def\Re{{\rm Re}\,}
\def\tg{\tilde g}
\def\tF{\tilde F}
\def\DF{\Delta F}
\def\p{\partial}
\def\quarter{{\textstyle{1\over4}}} 
\def\pie1{8\pi^2}
\def\SKIP#1{}
\def\etal{{\it et al,\/}\ }
\def\dmu#1{{d{#1}\over d\mu}}
\def\dtl{\delta_{\rm two-loop}}

\Title{
\vbox{\hbox{SSCL--Preprint--496}
\hbox{WIS--93/61/JULY--PH}}}%
{\vbox{\centerline{Light Threshold Effects in}
\centerline{ Supersymmetric Grand Unified Theories.}}}

\vskip-2in
Submitted to Nuclear Physics B.\ for publication
\bigfirst{\vskip1.6in}{\vskip1.4in}

\centerline{Alon E. Faraggi\footnote{$^\ddagger$}{
fhalon@weizmann. Address after September 1 1993, School of Natural Sciences,
Institute for Advanced Study, Olden Lane, Princeton NJ, 08540.}}
\centerline{Department of Physics, Weizmann Institute of Science}
\centerline{Rehovot 76100, Israel}
\medskip\centerline{and}
\medskip
\centerline{Benjam\'\i n Grinstein\footnote{$^\dagger$}%
{grinstein@hbar.ssc.gov, @sscvx1.bitnet, Address only good until
February 9 1994, unknown thereafter}}
\centerline{Superconducting Super Collider Laboratory}
\centerline{2550 Beckleymeade Ave, Dallas, Texas 75237, USA}

\bigfirst{\vskip .3in}{\vskip0.15in}

Supersymmetric Grand Unified Theories have a rich spectrum of
particles barely heavier than the intermediate vector bosons. As their
non-supersymmetric counterparts, they lead to many relations among low
energy observables. But the precise form of the predictions is
modified by the extended spectrum. If the masses of these new
particles are comparable to $M_Z$, the standard computation of their
effect becomes inaccurate. We present a detail discussion of the
correct procedure, and carry out the relevant computations to one loop
order.  The procedure we propose has the advantage over other existing
methods that two-loop running of gauge couplings can be incorporated
readily and consistently. Attention is paid to the special treatment
that the top and Higgs particles must receive. The size of the effect
is explored for a range of parameters in the minimal supersymmetric
$SU(5)$ grand-unified theory with radiative breaking. It is found that
the naive (leading-log) computation can be fairly inaccurate.

\Date{August 1993; REVISED November 1993}

\nref\adol{The ALEPH collaboration, \pl{B284}{1992}{163}\semi
The L3 collaboration, \pl{B284}{1992}{471}\semi
The DELPHI collaboration, Z. Phys. C {\bf 54} (1992) 55\semi
The OPAL collaboration, Z. Phys. C {\bf 55} (1992) 1\semi
The LEP collaborations, ALEPH, DELPHI, L3, OPAL, \pl{B276}{1992}{247}\semi
The ALEPH collaboration, Z. Phys. C {\bf 53} (1992) 1.}
\nref\ama{U. Amaldi \etal \prd{36}{1987}{1385}\semi
P. Langacker, Proc. PASCOS--90 Symposium, Eds. P. Nath and
S. Reucroft (World Scientific, Singapore, 1990)\semi
 P. Langacker and M. Luo, \prd{44}{1991}{817}\semi
J. Ellis, S. Kelley and D. V. Nanopoulos, \pl{B249}{1990}{441}\semi
\pl{B260}{1991}{131}\semi
U. Amaldi, W. de Boer and H. F{\"u}stenau, \pl{B260}{1991}{447}\semi
 H. Arason \etal \prd{46}{1992}{3945}\semi
F. Anselmo, L. Cifarelli, A. Peterman and A. Zichichi, \nuvc{105}{1992}{1179}.}
\nref\hall{R. Barbieri and L.J. Hall, \prl{68}{1992}{752}.}
\nref\fgm{A. E. Faraggi, B. Grinstein and S. Meshkov,
\prd{47}{1993}{5018}\semi
	 K. Hagiwara and Y. Yamada, \prl{70}{1993}{709}.}
\nref\ross{G. G. Ross and R. G. Roberts, \np{B377}{1992}{571}. See
also R. G. Roberts and L. Roszkowski, \pl{309B}{1993}{329}}

\nref\plnp{P. Langacker and N. Polonsky, \pr{D47}{1993}{4028}.}

\nref\ekn{J.~Ellis, S.~Kelley and D.~V.~Nanopoulos,
		\np{B373}{1992}{55}; \pl{B287}{1992}{95}. See also
Y.~Yamada, Z. Phys. C {\bf 60} (1993) 83.}

\nref\lynno{G. Passarino and M. Veltman, \np{B160}{1979}{151}\semi
A. Sirlin, \pr{D22}{1980}{285}\semi
A. Sirlin and W. J. Marciano, \np{B189}{1981}{442}\semi
M. Consoli, \np{B160}{1979}{208}\semi
F. Antonelli, M. Consoli and G. Corbo, \pl{91B}{1980}{90}\semi
M. Veltman, \pl{91B}{1980}{95}\semi
M. Lemoine and M. Veltman, \np{B164}{1980}{445}\semi
M. Green and M. Veltman, \np{B169}{1980}{137}; {\it erratum}
\ibid{B175}{1980}{547}\semi
M. Bohm and W. Hollik, \np{B204}{1982}{45}; {\it idem,\/}
\pl{139B}{1984}{213}\semi
M. B. Einhorn, D.R.T. Jones and M. Veltman, \np{B191}{1981}{146}\semi
W. J. Marciano and A. Sirlin,
\pr{D27}{1983}{552}; \ibid{D29}{1984}{75}; {\it ibid} p.945; {\it erratum,}
\ibid{D31}{1985}{213}\semi
W. J. Marciano and A. Sirlin, in {\it Brookhaven 1981, Proceedings,
Isabelle,}  Vol. 1, 289-302, 1981\semi
B. W. Lynn, M. E. Peskin and R. G. Stuart, in {\it Physics at LEP},
Vol.~1, J.~Ellis and R.~Peccei, eds., CERN~86-02, February 1986\semi
W. Hollik, Z.~Phys. \blankref{C37}{1988}{569}\semi
D. Kennedy and B.W. Lynn, \np{B322}{1989}{1}.}

\nref\lynnt{B. W. Lynn, SLAC-PUB-3358, June 1984 (unpublished)\semi
K. H. G. Schwarzer, OXFORD-TP 40/84, September 1984 (unpublished).}

\nref\lynpol{B. W. Lynn, in {\it Polarization at LEP,\/} Vol.~1,
G.~Alexander \etal eds., CERN~88-06, September~1988}

\nref\ghs{A. Giveon, L.J. Hall and U. Sarid, \pl{B271}{1991}{138}.}

\nref\bbo{See, for example, M. Carena, S. Pokorski and C. E. M. Wagner,
\np{B406}{1993}{59}\ and
V. Barger, M.S. Berger and P. Ohmann, \prd{47}{1993}{1093}, and
references thereof.}

\nref\martin{S. Martin and P.  Ramond, Sparticle Spectrum Constraints,
NUB-3067-93TH, UFIFT-HEP-93-16, SSCL-Preprint-439, June 1993 (unpublished).}

\nref\groups{J. Ellis and F. Zwirner, \np{B338}{1990}{317}\semi S.
Kelley, J. Lopez, H. Pois, D.V.  Nanopoulos and K. Yuan,
\pl{B273}{1991}{423}\semi M. Drees and M.M.  Nojiri,
\np{B369}{1992}{54}\semi P. Nath and R. Arnowitt,
\prl{69}{1992}{725}.} \nref\gpw{B. Grinstein, J. Polchinski and M. B.
Wise, \pl{B130}{1983}{285}}

\nref\okada{Y. Okada, M. Yamaguchi and T.  Yanagida, Prog. Theor.
Phys. Lett. 85 (1991) 1\semi J. Ellis, G.  Ridolfi and F. Zwirner,
\pl{B257}{1991}{83}\semi H. E. Haber and R.  Hempfling,
\prl{66}{1991}{1815}.}

\nref\velt{M. Veltman, \np{B123}{1977}{89}\semi
	M.S. Chanowitz, M.A. Furman and I. Hinchliffe, \pl{B78}{1978}{285} and
  \np{B153}{1979}{402}.}

\nref\hrs{L.J. Hall, R. Rattazzi and U. Sarid, The Top Quark Mass in
Supersymmetric $SO(10)$ Unification, LBL-33997, UCB-PTH-93/15,
hep-ph/9306309 (unpublished).}

\newsec{Introduction}
Grand Unified Theories possess many attractive features.
They simplify enormously the description of matter content of the
standard model by reducing the number of irreducible representations of
the gauge group needed to account for all observed particles. They
explain electric charge quantization. They bring all
non-gravitational interactions under a common umbrella treatment.
Yet, the minimal standard model extension to an SU(5) GUT is all but
ruled out: not only does proton not decay at the predicted rate, precise
measurements\adol\ of the gauge couplings of strong and electroweak
interactions have led to the observation that at no scale do they unify.

This rather grim observation is, nevertheless, not generic of GUT
theories, and it seems premature to dismiss the class of theories on
the basis of the failure of the minimal version. Remarkably, the
minimal supersymmetric extension of the minimal SU(5) GUT is free of
the aforementioned problems. That the three gauge couplings unify\ama\
in that case is not a surprise: an additional parameter, $\MS$, is
introduced which can be chosen so that couplings do unify. What is
remarkable is that unification occurs at a scale $\MG$ high enough
that proton stability is not in conflict with observation, yet bellow
the Planck scale, making it plausible that calculations with presently
understood techniques are sensible. Moreover, the additional
parameter, $\MS$, which describes the scale at which the running of
the gauge couplings changes from what is dictated by the standard
model to what is dictated by its supersymmetric extension, comes out
rather small, just about equal to the scale of electroweak symmetry
breaking. If one interprets $\MS$ as some average mass of the
supersymmetric partners of standard particles, the inescapable
conclusion is that a rich spectrum of new elementary particles is
awaiting discovery in the few hundred GeV range (well accessible to
planned next generation hadron colliders).

One ought not to rush into conclusions, though. Barbieri and Hall\hall\ have
pointed out that the standard analysis of running of gauge couplings
assumes all the superheavy particles (those with masses naturally of
order of $\MG$) are degenerate. The value of $\MS$, they argue, can be
shifted if one drops this assumption and includes the effects of the
non-degenerate superheavy thresholds. On closer examination, it has been
noticed\fgm\ that in the minimal SUSY SU(5) model the dependence of $\MS$ on
superheavy thresholds is rather weak.

For example, if $\as(M_Z)=0.125$%
\SKIP{with small errors\foot{ A literature
search shows that its value, as reported by LEP experiments over the
last two years, has wandered over few times the quoted error bars.}},
the value of $\MS$ exclusive of superheavy threshold effects comes out
to be well below the electroweak scale.  Superheavy thresholds can
push $\MS$ back up. But to obtain $\MS\ge M_Z$ one must have that the
superheavy supermultiplets that transform as $(8,1)_0 + (1,3)_0$ under
the gauge group $SU(3)\times SU(2)\times U(1)$ acquire masses some
eight orders of magnitude below $\MG$. Worse yet, to keep $\MG$ well
below the Planck scale one must further complicate the spectrum of
superheavies, splitting the masses of these two multiplets by two
orders of magnitude. The simple compelling picture of a dessert
between the electroweak and GUT scales is lost. Instead one must
populate the dessert with particles at two new intermediate scales.
For the remaining of this work we neglect these heavy threshold
effects.

In addition to superheavy threshold effects, one must consider the
effects of light thresholds (non-degeneracy of the SUSY particles with
masses naturally of order the electroweak scale). Ross and
Roberts\ross\ argue that because in most models the average mass of
the colored superpartners is larger than that of the uncolored ones, a
simple parameterization in terms of a single threshold at $\MS$ may be
misleading. Langacker and Polonski\plnp\ point out that one need not
interpret $\MS$ as some sort of average mass for the spectrum of SUSY
particles.  The spectrum of the model is not expected to be simple.
One can imagine accounting for the complicated spectrum in some
accurate fashion, and then finding a scale $\MS$ which would mimic the
effects of a complicated spectrum.  Only this scale does not have a
clear physical meaning. In fact, they find it possible to construct
examples of rather heavy SUSY spectra (compared with $M_Z$), yet
having $\MS \ll M_Z$.

The standard analyses of light threshold effects consists of modifying
the evolution equations for the gauge couplings at each consecutive
threshold\ekn. More specifically, starting from unified gauge
couplings at $\MG$, $\a_1(\MG)=\a_2(\MG)=\a_3(\MG)$, one uses the
renormalization group equation (RGE) to evolve these couplings down to
the first threshold. The evolution equations are modified by reducing
the number of active degrees freedom that affect the running, and the
couplings are run further down to the next threshold, and so on.  In
what follows we shall refer to this method as the `run-and-match' or
the `naive' method. This method is justified, and expected to be a
good approximation, provided all of the SUSY particles are much
heavier than the $Z$-boson. Also, this kind of analyses by necessity
assumes unbroken electroweak symmetry, which again is reasonable at
mass scales well above $M_Z$.

In this paper we address the issue of how to compute these light
threshold effects when the SUSY particles are not necessarily much
heavier than the $Z$-boson. The motivation for this comes not just
from the discussion just given, but from the recent observation that
phenomenologically viable models that incorporate SUSY SU(5) and
radiative electroweak breaking invariably contain many particles with
masses barely exceeding $M_Z$. Moreover, we will craft the method of
computation so that two-loop running effects are incorporated
consistently.

We will first formulate carefully the question we want to address.
This is done in sections~2 and~3 where we give the fundamental
equations that allow us to relate the measurements of gauge couplings
as reported by, say, LEP/SLC experiments to the gauge couplings of the
SUSY model.  The important discussion of the effects of
two-loop running can be found in section~4. The difference between a
calculation of threshold effects based on effective field theories
versus calculations based on, say, the so called ``star scheme'' is
discussed there.  In section~5 we develop general formulae for
threshold effects modifications to relations among parameters dictated
by grand unified theories. In section~6 we exhibit the result of one
loop computations of the quantities that go into the general formulae
of the preceding section. Special care must be taken in dealing with
the Higgs sector and the top quark, and this is addressed in
section~7. In preparation for the discussion and presentation of our
numerical analysis of section~9, we review the minimal supergravity
$SU(5)$ model in section~8.  Our conclusions can be found in
section~10.

There is a long history of calculations of radiative corrections in
and beyond the electroweak theory\lynno. Calculations in
supersymmetric theories date back to refs.~\lynnt. No doubt the trivial
calculations of one loop diagrams in this paper can be found
elsewhere.  The emphasis in this paper is on the treatment of light
threshold effects in grand-unified theories, and in particular in
SUSY-GUTs. It may not be obvious to the reader but, as we will see
(sect.~3), the computations involved coincide with those of radiative
corrections to electroweak parameters. Thus, for example, while
ref.~\lynpol\ presents a detailed discussion of radiative corrections
to electroweak parameters in SUSY extensions of the standard model
(see table 1 and figs.~39--45 of that work), the same work treats the
case of grand-unified SUSY theories as containing a single common mass
threshold at a scale `$\mu$' (see figs.~47--49 there). We believe ours
is therefore the most complete treatment of the light thresholds in
SUSY GUTs to date.  Moreover, we have failed to find as detailed a
presentation of the numerical size of the effects for the minimal SUSY
$SU(5)$ GUT theory with radiative breaking as we give here ({\it
cf,\/} sect.~8).

\newsec{Threshold Corrections}
Let us define what is meant by threshold corrections in this paper.
The gauge sector of the standard model of electroweak and strong
interactions has four independent parameters: the three gauge
couplings\foot{We assume, for definiteness, that dimensional
regularization is used and that lagrangian parameters are defined
through minimal subtraction.} $g_{1,2,3}$ and the vacuum expectation
value $v$. These parameters are not all independent in a unified
theory. For example, in minimal SUSY SU(5) GUT one has
\eqn\sintree{
\st = {1\over5} +{7\over15}{\ae\over\as}~,
}
where $\st=g_1^2/(g_1^2+g_2^2)$, $\ae=g_1^2g_2^2/4\pi(g_1^2+g_2^2)$
and $\as=g_3^2/4\pi$ are all defined in terms of the running coupling
constants, and the relation holds as a function of the renormalization
scale $\mu$. This relation holds in the leading log approximation
(corrections are suppressed by $1/\log(\MG/\mu)$). It assumes a
spectrum of particles much lighter than $\MG$, the ``light
particles'', consisting of the particles in the standard
model with two higgs doublets, and adding the superpartner of
each field.

To test whether this relation holds one must carefully define the
terms that enter into it. It would seem appropriate to cast this
relation in terms of physical observables. For example, one may write
\eqn\sintestone{
{1\over2}\left(1-\sqrt{1-{4\pi\ae\over\sqrt2G_\mu M_Z^2}}\right)=
 {1\over5} +{7\over15}{\ae\over\as}
}
or
\eqn\sintesttwo{
1-M_W^2/M_Z^2 = {1\over5} +{7\over15}{\ae\over\as}~. }
These are not equivalent, and cannot both hold. The discrepancy comes
from the difference, beyond tree level, between two quantities that
agree at tree level. Given a precise definition of the four observables
$G_\mu$, $M_Z$, $\ae$ and $\as$, the equations can be corrected to account for
the difference:
\eqna\sintestcorrctd
$$
\eqalignno{ {1\over2}\left(1-\sqrt{1-{4\pi\ae\over\sqrt2G_\mu
M_Z^2}}\right)&= {1\over5} +{7\over15}{\ae\over\as}+\delta_1
&\sintestcorrctd a\cr
1-M_W^2/M_Z^2 &= {1\over5}
+{7\over15}{\ae\over\as}+\delta_2 &\sintestcorrctd b\cr } $$
Contributions to $\delta_{1,2}$ fall into two distinct classes: those
that are there in the standard model, and those that arise from
the virtual effects of new particles beyond the standard model's. It is
the latter that we would like to focus on and which give rise to what
we call ``light threshold effects''. The idea is that one need not
recompute the effect of the standard model particles every time an
analysis like this is performed. Moreover, one needs not grope with the
issue of which physical observables to use to represent $\st$.
Instead, one may perform a standard model based analysis to extract
the values of the couplings $g_{1,2,3}$ and $v$. If there is an
underlying SUSY SU(5) GUT, the couplings extracted from an analysis based on
the full spectrum are then expected to satisfy a relation like in eq.~\sintree.
If only the standard model's spectrum is used in the extraction of
the couplings, there will be a small correction
\eqn\dlightdefd{
\st = {1\over5}+{7\over15}{\ae\over\as}+\dl~,
 }
where $\dl$ is precisely what we intend to mean by ``light threshold
corrections''. Eq.~\dlightdefd\ is used in ref.~\ekn\ to implicitly
define light threshold effects. The arguments leading to $\dl$ there
are different and simpler, given the nature of the approximations in
that paper. Here we intend to generalize the results of ref.~\ekn, and
we will find it useful to check that our results for $\dl$ reduce to
those of ref.~\ekn\ in the appropriate limit.

\newsec{Generalities}

Gauge coupling constants are not physical observables. When an
experiment reports on the value of a gauge coupling, there is an
implicit translation of some observables into these theoretical
constructs. This is not to say that the values of gauge couplings lack
in importance. On the contrary, given a well specified definition of
these couplings, they encode concisely the results of measurements.

When an experimental result is analyzed to extract the value of gauge
coupling constants, the actual value obtained depends in detail on the
theoretical assumptions. Given two different models with the same
gauge group but different particle content, as is the case of the
standard model and its supersymmetric extension, the extracted values
of gauge couplings from the same set of observables is a priori
different for the two models. In principle one could analyze the
observables directly under the different set of assumptions, \ie,
different models, and thus extract the values of gauge coupling
constants appropriate to those assumptions. In practice however, it is
impossible to reanalyze the experiment for each new set of
assumptions.

The results of experiments are therefore most often given in terms of
an analysis based on the standard model. This makes sense. The
standard model is the most concise model of elementary interactions
consistent with all present observations. Clearly what we need is a
means for translating experimentally determined values of gauge
couplings in the standard model into gauge couplings in extensions of
the standard model.

It is possible to have such a translation. An obvious case is one in
which the model at hand is an extension of the standard model that has
only very heavy particles with hard masses. By this we mean that their
masses are much larger than the energies at which the experiments are
carried out, and that the masses are symmetric under the electroweak
$SU(2)\times U(1)$, up to small corrections. Then the decoupling
theorem guarantees that the physics at experimental energies is
described by an effective Lagrangian which corresponds precisely to
the standard model. The effective coupling constants of this effective
theory are numerically equal to  the standard model's. The gauge
couplings of the underlying theory are easily related to those of the
effective theory, by matching conditions at the scale of the heavy
particles. Non-supersymmetric GUTs furnish an example. The underlying
coupling constant, $\ag$, can be derived by matching at $\MG$, so one
must determine experimentally the electroweak and strong couplings at
a low scale, and then run them up to $\MG$ where the effective theory
can be matched to the full underlying theory.

In fact, this method can also be used even when the heavy particles
are only moderately heavier than the energy scale of experiments. This
is because this method accounts for all effects that are logarithmic
in the ratio of the experimental scale $E$ and the heavy mass $M$,
$\ln(E/M)$; corrections are suppressed by powers of $E/M$, or even of
$(E/M)^2$. Therefore it is quite sensible to use this method to
analyze SUSY GUTs, especially if one is interested in a spectrum of
heavy particles in the TeV range.  In fact, this is precisely the way
in which the low threshold effects have been accounted for by many in
the past\refs{\ross--\ekn,\ghs}.  We refer to this method as the
`run-and-match' or `naive' calculation.

But one may ask what the proper procedure is, and even whether one
exists, for the case in which the heavy particles are only barely
heavier than the experimental scale, \ie, $E/M$ is only slightly
smaller than unity. After all, many recent analysis of SUSY GUTS show
that models typically contain several particles in the few
hundred GeV range. And not only is $E/M$ only slightly smaller than~1;
since the multiplicity is high, one expects to find large coefficients
in front of the order $(E/M)$ correction.

It turns out it is easy to construct the proper procedure. We will
assume that particles are heavy enough that they cannot be directly
produced at present (else, the theory should be ruled out or confirmed
experimentally in short order!). This is important because it implies
immediately that the only observables available are those of the
standard model. In other words, we can consider, say, cross sections
or decay widths of standard model particles for comparison with
experiment, and not consider any novel processes of the extended
theory. Let $O_i$, $i=1,2,3,\ldots$ stand for a collection of
observables. For example, this can be taken to be the forward-backward
asymmetry at the peak of the $Z$-resonance, the differential cross
section $d\sigma(e^+e^-\to \mu^+\mu^-)/d\cos\theta$ at a collection of
specific center of mass energies and angles, and similar cross
sections into other leptons or quarks. The standard model gives
expressions $F_i$ for these observables in terms of its parameters
$g_k$,
\eqn\observea{
O_i=F_i(g_k)~.
}
Now, any extension of the standard model will also give expressions
$\tF_i$ for the same observables in term of its own parameters.
Among these parameters are those which have a direct correspondence to
those of the standard model, $\tg_k$, like gauge couplings and
particle masses. In addition there are new parameters $e_n$
characterizing the new physics, like new particle masses. So one has,
\eqn\observeb{
O_i=\tF_i(\tg_k, e_n)~.
}
Although conceptually the same, one must differentiate between the
couplings $g_k$ and $\tg_k$. In fact, it is this difference that
we are trying to calculate for the case of gauge couplings! In other
words, the problem at hand is to calculate $\tg_k$ from knowledge
of $g_k$, given an assumption on the values of the parameters $e_n$.

Our task is to invert eqs~\observea--\observeb\ for $\tg_k$ in
terms of $g_k$ (and, implicitly, $e_n$). Now, the additional degrees
of freedom enter into the expressions for observables as  heavy virtual
particles. Therefore, their effects are small. We can write
\eqn\gtog{
\tg_k = g_k + \dg_k~,
}
where $\dg_k$ are small one loop order corrections which are
expressed as functions of $g_k$ (and, implicitly, $e_n$). It turns
out to be convenient to write the functions $\tF$ as a sum of
two pieces, one identical to the standard model's, plus the rest:
\eqn\ftildesplit{
\tF_i(\tg_k, e_n) = F_i(\tg_k) +\DF_i(\tg_k, e_n)
}
This is convenient because $\DF$ can be regarded as a small quantity,
the same order as $\dg$. This means also that we are assuming the same
renormalization scheme and gauge choice in both models. Combining
equations and keeping only leading order in small quantities, one
obtains
\eqn\mastereq{
\sum_l\dg_l{\p F_i\over\p g_l}(g_k) + \DF_i(g_k,e_n) = 0~.
}
Equation \mastereq\ is a set of simultaneous equations for $\dg_k$. In
the first term, the derivative ${\p F_i\over\p g_l}$ should be
evaluated at tree level, because the coefficient $\dg_l$ is already of
one loop order. This is a welcome simplification: we only need to
compute the one loop effects of the new particles, because the standard
ones have already been included in the experimental determination of
the couplings $g_k$.

It is important to note that this discussion depends crucially on
approximate decoupling. Were the new particles not heavy enough,
the dependence of observables on kinematic variables (like $s$
and $t$ in cross sections) would differ from one model to the other.
To the extent that deviations from the standard model cannot be
inferred from kinematic dependence of observables one can neglect
these kinematic effects in $\DF$. Although this observation seems
rather innocuous, it is of practical importance. We need it to ensure
compatibility of expressions \observea\ and~\observeb.

Now, it would seem the program of determining the threshold
corrections $\dg_k$ is complete: calculate the functions $F_i$ at tree
level, and the virtual effects of the new particles at one loop
$\DF_i$, and plug into eq.~\mastereq. Then solve for $\dg_k$.
Still, this program involves a choice of specific observables,
labeled by $i$. Recall that  we would like to be able to connect the
$\tg_k$ to the $g_k$ without detailed knowledge of the experimental
data.

The Feynman diagrams that contribute to the functions $\DF_i$ can be
classed in two groups: corrections to standard particle propagators
and vertex or box corrections. The former are process independent,
\ie, they enter in a universal form into the relation between $\tg$
and $g$. The latter depend on each process, but are often smaller than
the former. We therefore drop them. In standard lingo, we retain only
oblique corrections. In any case this kind of truncation is implicitly
done in the run-and-match approach. Our analysis is intended as an
improvement on that approach, but it is clear that it is incomplete in
this regard. (It should also be clear, from the above discussion, that
short of a complete reanalysis of experimental data, this is the best
one can do).

\newsec{ Two Loop Running}
The relation between coupling constants in eq.~\sintree,
$$\st = {1\over5} +{7\over15}{\ae\over\as}~,
\eqno\sintree$$
follows from the solution to the renormalization group equations at
one-loop order
\eqn\rgeoneloop{
\mu\dmu{g_i} = b^{(i)}_0 g_i^3~,
}
namely
\eqn\rgesolsoneloop{
{1\over g(\mu)^2}-{1\over g(\mu_0)^2}=b_0\ln{\mu_0^2\over\mu^2}~.
}
At two-loops the renormalization group equations are difficult to
solve. They are a set of coupled equations. Neglecting
yukawa couplings one has
\eqn\rgetwoloop{
\mu\dmu{g_i} = b^{(i)}_0 g_i^3+\sum_j b^{(ij)}_1 g_i^3 g_j^2~.
}
How does this modify the relation between gauge couplings in
eq.~\sintree{}? In principle there is an analytic expression that
expresses the new relation between the couplings. In practice one
recognizes that, because the gauge couplings are small, the
corrections to the relation in eq.~\sintree\ are correspondingly small.
So one may write
\eqn\sintwoloop{
 \st = {1\over5} +{7\over15}{\ae\over\as} +\dtl~.
}
It is customary to calculate the correction $\dtl$ by solving the
renormalization groups eqs.~\rgetwoloop\ numerically and comparing to
the solution of the one-loop version, eq.~\rgesolsoneloop.

There are two important points that must be stressed:
\item{(i)} The combined correction from light threshold and two-loop
running effects is computed by simply adding together the
corresponding corrections, $\dl$ and $\dtl$; and
\item{(ii)} Both corrections ($\dl$ and $\dtl$) are formally of the same
order, namely $\CO(\a_i)$.

The first of these points is not only obvious but holds to arbitrary
order. The low energy effective theory has gauge couplings which
satisfy renormalization group equations. In principle, these may be
solved to arbitrary accuracy. In the effective theory one can then
make predictions of physical observables, and carry out the
computations to the same accuracy as the running. Effective field
theories are particularly good at making this sort of logic, and the
corresponding calculation, transparent.

Consider, in contrast, the prediction using the ``star''
scheme\lynpol. One is instructed to compute star coupling
constants $g_{i\ast}(q^2)$ in terms of the bare coupling constants
$g_{i~{\rm bare}}$,
\eqn\gstar{
{1\over g^2_{i\ast}(q^2)}-{1\over g^2_{i~{\rm bare}}} = \Pi_i(q^2)
}
where the $\Pi_i(q^2)$ are appropriately chosen two-point functions,
directly related to self-energies of gauge-particles. It is easy to
see how the star and effective field theory approaches give the same
answers at one loop. For example, the right sides of eq.~\gstar\ is
dominated by a large logarithm when $q^2$ is very small compared to
the cut-off, $\Lambda^2$, and this logarithm precisely corresponds to
the one in eq.~\rgesolsoneloop\ if we make the correspondence
$|q^2/\Lambda^2|\to\mu^2/\mu_0^2$.

In the star scheme a two loop contribution to $\Pi_i(q^2)$ is
necessarily of order $g_i^2$. Therefore, it gives a correction
$\delta g_i^2/g_i^2$ of order $g_i^2g_j^2$. This is in contrast with
the correction $\dtl$, which gives $\delta g_i^2/g_i^2\sim g_j^2$.
To see this, we may  solve a simpler equation that does the power counting
correctly. Take, for example, pure QCD, which corresponds to
neglecting the couplings $g_{1,2}$ in the running of $g_3$ in
eq.~\rgetwoloop,
\eqn\rgeqcd{
\mu\dmu{g_3}=b_0g_3^3 + b_1g_3^5~.
}
The solution is well known, and has a simple expression if one retains
only leading terms in $g_3$,
\eqn\rgeqcdsol{
{1\over g_3(\mu)^2}-{1\over g_3(\mu_0)^2}
=b_0\ln{\mu_0^2\over\mu^2}
+{b_1\over b_0}\ln{g_3(\mu_0)^2\over g_3(\mu)^2} ~.
}
The second term on the right side cannot result from the two loop
computation of the two-point function $\Pi$ in eq.~\gstar. It is not
clear to us how the star scheme can incorporate the two-loop running,
if at all.

That $\dl$ and $\dtl$ are of the same order is readily seen by explicit
computation. Indeed, an expression for $\dl$ appears in the next
section, and is explicitly of order $\a$. And comparing
eqs.~\rgesolsoneloop\ and~\rgeqcdsol\ one infers the correction to the
running coupling $\delta g^2$, with $\delta(1/g^2)=-\delta g^2/g^4\sim
\CO(g^0)$, or $\delta g^4\sim g^4$, which leads to $\dtl\sim\a$.

It is worth noting that formally the heavy-threshold effects
$\delta_{\rm heavy}$ are of the same order as $\dl$ and $\dtl$.\hall\fgm

Numerical analysis of the effects of two loop running (including
yukawa couplings) has been extensive \bbo. Our numerical results do not
include the effects of two-loop running. Our emphasis is on the light
threshold effects and how they depend on the spectrum. The effects of
two-loop running are independent of the details of the spectrum. The
interested reader may incorporate the two-loop running effect simply
by adding an overall constant (\ie, spectrum independent), $\dtl$, to
our predictions for $\st$.

\newsec{Derivation of Master Formulas.}
We are ready to proceed with explicit calculations. Although the
expression for the threshold corrections $\dg_k$ will be  process
independent, it is necessary to choose particular (gedanken) process
to obtain it.

We first consider a particularly simple example, $\mu\bar\nu_\mu \to
e\bar\nu_e$ scattering. We hope this will serve to illustrate the
general approach with a minimum of unnecessary complications. We take
the observable $O$ as the cross section (at some kinematic point),
divided by the phase space, \ie, the square-modulus of
the invariant amplitude,
\eqn\ampli{ O = F(g) = \sum_{\rm helicities} |{\cal A}|^2
}
At tree level this is given by the Feynman diagram in~\fig\one{Tree
level Feynman diagram for the scattering amplitude for
 $\mu\bar\nu_\mu \to e\bar\nu_e$ scattering.}

Now, we really only need to extract the dependence on couplings,
without paying careful attention to the kinematic factors. Therefore
we write
\eqn\reducedamp{
\sum_{\rm helicities} |{\cal A}|^2 =
\left({g_2^2\over s-g_2^2v^2}\right)^2
\sum_{\rm helicities} |\tilde{\cal A}|^2 ~.
}
Here $s$ is the standard kinematic variable, $g_2$ is the $SU(2)$
coupling constant and $v$ is the electroweak breaking vacuum
expectation value. Clearly $\tilde{\cal A}$ is just a product of
spinors and gamma matrices.

The one loop virtual effects of new heavy particles includes vertex,
box and propagator corrections; these are shown in \fig\two{Feynman
diagrams for one loop virtual effects of new heavy particles.}. As
discussed previously, we will neglect the vertex corrections. The
propagator corrections are conveniently expressed in terms of the
self-energy (defined as the 1PI amputated two point function for the
charged vector boson): \eqn\selfenrgy{
\Pi_{W\mu\nu}(q) =(q^2g_{\mu\nu}-q_\mu q_\nu)\Pi_W^{(T)} +
q_\mu q_\nu\Pi_W^{(L)}
}
In terms of these the $W$-boson propagator in Feynman-'tHooft gauge is
\eqn\Wprop{
-i{g_{\mu\nu}-q_\mu q_\nu/q^2\over q^2 -g_2^2v^2-q^2\Pi_W^{(T)}} -
i {q_\mu q_\nu/q^2\over q^2 -g_2^2v^2-q^2\Pi_W^{(L)}}
}

The function $\DF$ that we need to compute is defined by
eq.~\ftildesplit, which tells us that we should take the difference
between $|{\cal A}|^2$ in the extended model and in the standard
model. The tree level amplitudes are equal in both models, so the
difference comes from the cross product of the tree level amplitude
and the one-loop amplitude:
\eqn\oneloopDF{
\DF=2g_2^4\,\Re {s\hat\Pi^{(T)}_W\over (s-g_2^2v^2)^3}
\sum_{\rm helicities} |\tilde{\cal A}|^2 ~.
}
Here by $\hat\Pi_W^{(T)}$ we mean that part of the self-energy which
is not already in the standard model calculation, \ie, which comes from
the new particles of the extended model running around in the loop.
Using eqs~\ampli--\oneloopDF\ in eq.~\mastereq, one obtains
\eqn\preresultgtwo{
\left[\dg_2 {\p\over\p g_2} + \dv {\p\over\p v}\right]
\left({g_2^2\over s-g_2^2v^2}\right)^2
+ {2g_2^4 s\over(s-g_2^2v^2)^3} \Re \hat\Pi_W^{(T)} =0~.
}
or, simplifying,
\eqn\gtwo{
{\dg_2^2\over g_2^2} + {g_2^2v^2\over s}{\dv^2\over v^2}
= -\Re \hat\Pi_W^{(T)}(s)
}

Similar expressions can be obtained for other couplings by considering
other processes. From neutrino scattering one obtains
\eqn\gonetwo{
{\dg_1^2 +\dg_2^2\over(g_1^2 + g_2^2)}
+ {(g_1^2+g_2^2)v^2\over s}{\dv^2\over v^2}
=            - \Re \hat\Pi_Z^{(T)}(s)~,
}
while from quark scattering
\eqn\gthree{
\dg_3^2 = -g_3^2 \Re \hat\Pi_g^{(T)}(M_Z^2)~.
}
Here\foot{In this section, the normalization of $g_1$ is the standard
model one. In the following sections we will switch, without warning,
to the normalization that is more appropriate for GUT theories. That
is, we will absorb a factor of $\sqrt{5/3}$ in~$g_1$.}
$g_1$ and $g_3$ stand for the gauge couplings for the $U(1)$
and $SU(3)$ groups, respectively, and $\Pi_Z$ and $\Pi_g$ are the
$Z$-boson and gluon self energies. Again, the hat over these
quantities reminds us to include only non-standard model
contributions.

At first sight these expressions appear nonsensical: the dependence on
the kinematic variable, $s$, is manifestly different on both sides of
the equation. Were high precision experiments performed over a wide
range of energies, one could either verify or rule out these effects.
In practice precision experiments are available only at very low
energies, $s\ll M_W^2$, or on resonance, $s\approx M_W^2$ or $M_Z^2$.
We will not assume that the functional form of $\hat\Pi_W^{(T)}(s)$ or
$\hat\Pi_Z^{(T)}(s)$ can be well approximated by truncating their
expansion to contain only a pole and constant terms. Although this
would match the functional form on both sides of eqs.~\gtwo\
and~\gonetwo, one of our objectives is to retain the full effects of
these self-energies. Thus, from eqs.~\gtwo\ and~\gonetwo\ one obtains a
set of four equations for three unknowns:
\eqna\simulteqs
$$\eqalignno{
{\dg_2^2\over g_2^2} + {\dv^2\over v^2}
&= -\Re \hat\Pi_W^{(T)}(M_W^2) &\simulteqs a \cr
{\dg_1^2 +\dg_2^2\over(g_1^2 + g_2^2)}
+ {\dv^2\over v^2}
&=            - \Re \hat\Pi_Z^{(T)}(M_Z^2)&\simulteqs b\cr
 {\dv^2\over v^2}
&= -\lim_{s\to0}{s\over M_W^2}\Re \hat\Pi_W^{(T)}(s)&\simulteqs c\cr
 {\dv^2\over v^2}
&= -\lim_{s\to0}{s\over M_Z^2}\Re \hat\Pi_Z^{(T)}(s)&\simulteqs d\cr
}
$$

Clearly, the last two equations are not generally mutually compatible.
It is easy to see that the difference between the right hand side of
eqs.~\simulteqs c\ and~\simulteqs d\ is a contribution too Veltman's
$\rho$ parameter. There are two ways of dealing with this issue. Since
there is no direct evidence for deviations from the standard model,
one may consider such contribution to $\rho$ negligible. This is in
keeping with the general philosophy of our method. Nevertheless, one
may object that the experimental determination of the $\rho$ parameter
is not at the same level of precision as, say, measurements of $M_Z$
or forward-backward assymmetries. Therefore, our theoretical results
for, say, $\st$ would inherently carry an error which could be much
larger than the precision with which it is measured. Alternatively,
one may introduce the $\rho$ parameter itself into the analysis. This
corresponds to replacing eqs.~\simulteqs b and~\simulteqs d\ by
$$ \eqalignno{
{\dg_1^2 +\dg_2^2\over(g_1^2 + g_2^2)}
+ {\dv^2\over v^2} + {\delta\rho\over\rho}
&=            - \Re \hat\Pi_Z^{(T)}(M_Z^2)&\simulteqs {b'}\cr
{\dv^2\over v^2}+ {\delta\rho\over\rho}
&= -\lim_{s\to0}{s\over M_Z^2}\Re \hat\Pi_Z^{(T)}(s)
		&\simulteqs {d'}\cr}
$$
Although $\rho$ is not a fundamental parameter of the standard model, it
is usually included in experimental fits.

We will investigate both approaches. One cannot choose one over the other as a
matter of principle: it is the way that the experimental fit is performed that
determines which analysis should be chosen. We present both for comparison.
 Clearly it is of interest to carry out the same type of
analysis with the extended set of six parameters of Golden and
Randall\ref\golden{M. Golden and L. Randall, \np{B361}{1991}{3}}\ and Altarelli
and Barbieri\ref\altbarb{G.  Altarelli and R. Barbieri, \pl{B253}{1991}{161}}\
--- adding not just $\rho$ to the standard model's. We will return to such
analysis in a forthcoming publication.

In what follows we will refer to the approaches that neglect and include the
contributions to $\rho$ as method--I and method--II, respectively.

We are now ready to compute threshold corrections $\dl$ as defined in
section~2, eq.~\dlightdefd.
\SKIP{
\eqn\dtheta{
\delta \st = \ct \,\Re [\hat\Pi_W^{(T)}(M_W^2) -
 \hat\Pi_Z^{(T)}(M_Z^2) ]
}
Here, the weak mixing angle is defined in terms of the gauge coupling
constants, $\st = g_1^2/(g_1^2+g_2^2)$.
Also, of specific interest to the SUSY-GUT theories is the light
thresholds correction to the leading-log relation
Expressing this relation in terms of the couplings as extracted from
experiment, that is, using the standard model couplings, one finds
corrections to the relation,
The correction $\dl$ was computed, using the run-and-match approach in
ref.~\ekn. In section~8 we will compare the results of that and our
calculation of $\dl$.
}
Using the results  above, one obtains in method--I
\bigfirst{\eqn\dlighti{
\dl^{(I)} = {\ae\over20\pi} \left[
{-1+2\st-5\sin^4\theta \over\sin^4\theta}\Phi_W
+{5\st-1\over\st}\Phi_W^0
+  {1\over\sin^4\theta} \Phi_Z +{7\over3} \Phi_g \right]
}}{\eqn\dlighti{
\dl^{(I)}\! =\! {\ae\over20\pi} \left[
{-1+2\st-5\sin^4\theta \over\sin^4\theta}\Phi_W
\!+\!{5\st-1\over\st}\Phi_W^0
\!+\!  {1\over\sin^4\theta} \Phi_Z\! +\!{7\over3} \Phi_g \right]
}}
where we have introduced the reduced self-energies
\eqna\phidefd
$$\eqalignno{
 \Re\hat\Pi_W^{(T)}(M_W^2) &= {\ae\over4\pi}{1\over\st}
		         \Phi_W &\phidefd a\cr
\lim_{s\to0}s \Re\hat\Pi_W^{(T)}(s) &= {\ae\over4\pi}{M_W^2\over\st}
		         \Phi_W^0 &\phidefd {b}\cr
 \Re\hat\Pi_Z^{(T)}(M_Z^2) &= {\ae\over4\pi}{1\over\st\ct}
           \Phi_Z &\phidefd c\cr
 \Re \hat\Pi_g^{(T)}(M_Z^2) &=  {\as\over4\pi}
          \Phi_g &\phidefd d\cr }
$$
These are functions of the masses of the virtual particles. When any
one such mass becomes large, the leading contribution  to these
functions is in the form of a logarithm of the ratio of the mass to
the $M_Z $ or $M_W$. Thus one recovers the form of the result of the
usual method, described earlier, of matching couplings and changing
evolution equations at thresholds. We have checked that our
expressions agree  with the standard run-and-match formulas when we
take the large mass, $SU(2)$ symmetric limit.

The threshold corrections  in method--II are
\eqn\dlightii{
\dl^{(II)} = {\ae\over20\pi} \left[
{-1+2\st-5\sin^4\theta \over\sin^4\theta}(\Phi_W-\Phi_W^0)
+  {1\over\sin^4\theta} (\Phi_Z-\Phi_Z^0) +{7\over3} \Phi_g \right]
}
where in addition to the definitions in \phidefd{}, we have introduced
\eqn\phizdefd{
\lim_{s\to0}s \Re\hat\Pi_Z^{(T)}(s) = {\ae\over4\pi}{M_Z^2\over\st\ct}
		         \Phi_Z^0 ~.
}

\newsec{Explicit Computations}
In the previous section we set up the stage for explicit computations.
In this section we present explicit results of one-loop computations.

Lets consider first the case of scalar particles contributions to
$\Phi_X$. The two required one-loop Feynman diagrams are shown in
\fig\scalardiags{ One-loop Feynman diagrams in the calculation of
threshold corrections: scalar particle contributions.}. The
calculation is performed using  dimensional regularization and minimal
subtraction. The result can be written succinctly in terms of the function
\eqn\ionedefd{
I_1(m_1,m_2,q)={m_1^2\over q^2}\ln{m_1^2\over \mu^2} +
	{m_2^2\over q^2}\ln{m_2^2\over \mu^2}-{1\over3}
	-2\int_0^1dx\,{M^2(x)\over q^2}\ln {M^2(x)\over \mu^2}~,
}
where
$$
M^2(x)\equiv m_2^2 x + m_1^2 (1-x) - q^2 x(1-x)
$$
and $\mu$ is the renormalization point. For a colored complex scalar
of mass $m$ in the $R$ representation of color $SU(3)$, with Casimir
invariant $C(R)$, one has
\eqn\Phigscalar{
\Phi_g= C(R) I_1(m,m,M_Z)~.
}

The all important effect of mixing enters the computation of $\Phi_Z$ and
$\Phi_W$. Writing the scalar field component of the neutral current as
\eqn\neucurrscalars{
g_{ij}~\phi^\ast_i \darr\partial_\mu \phi_j~,
}
and the corresponding scalar masses as $m_{i,j}$, one obtains the $Z$
self-energy \eqn\PhiZscalarmix{
\Phi_Z =|g_{ij}|^2 I_1(m_i,m_j,M_Z)~.
}
The coupling $g_{ij}$ includes the usual group theoretic factors and gauge
couplings, and the angles from the mixing matrix. Consider, for a definite
example, a scalar of mass $m$ which does not mix, that is, for which the mass
and weak eigenstates agree. We only need consider scalars in representations
which are singlets or doublets under weak-$SU(2)$. Let $\tau^3$ stand for the
third component of weak-isospin and $Q$ for electric charge. One obtains,
\eqn\PhiZscalar{
\Phi_Z = (\tau^3-\st Q)^2 I_1(m,m,M_Z)~.
}
The expression for $\Phi_W$ is entirely analogous to that of $\Phi_Z$ in
eqn.~\neucurrscalars. Lets consider the case of no mixing. $\Phi_W=0$ for
weak-singlets, while if the members of the doublet have masses $m_1$ and $m_2$
\eqn\PhiWscalar{
\Phi_W=\half I_1(m_1,m_2,M_W)~.
}
Also,
\eqnn\PhiWoscalar
$$\eqalignno{
\Phi_W^0 &=\lim_{q\to0}{q^2\over2 M_W^2} I_1(m_1,m_2,q) & \cr
&={1\over4M_W^2}\left[m_1^2+m_2^2 +{2m_1^2m_2^2\over
m_2^2-m_1^2}\ln{m_1^2\over m_2^2}\right]~. & \PhiWoscalar \cr
}
$$

Next we turn to the case of fermions. There is only one Feynman
diagram contribution to $\Phi_X$ at one-loop; see
\fig\fermiondiag{One-loop Feynman diagram in the calculation of
threshold corrections: fermion contributions.}. In the case of
fermions, three functions are needed to describe the results:
\eqna\itwothreedefd
$$
\eqalignno{
I_2(m_1,m_2,q) &=8\int_0^1dx\,x(1-x)\ln M^2(x)/\mu^2
					&\itwothreedefd a\cr
I_3(m_1,m_2,q) &=-{4\over q^2}\int_0^1dx\,[m_2^2x+m_1^2(1-x)]\ln M^2(x)/\mu^2
					&\itwothreedefd b\cr
I_4(m_1,m_2,q) &={4m_1m_2\over q^2}\int_0^1dx\,\ln M^2(x)/\mu^2
					&\itwothreedefd c\cr
}
$$
In terms of these, one has for colored fermions
\eqn\Phigfermion{
\Phi_g=C(R)I_2(m,m,M_Z)~.
}
We will be generally concerned with right handed, as well as left
handed,  weak-doublets. Therefore it is useful to introduce some
additional notation, allowing for the possibility of mixing in the
neutral current:
\eqn\neutralcurr{
{e\over2\sin\theta\cos\theta}
 	\bar\psi_i(g^L_{ij}(1-\gf)+g^R_{ij}(1+\gf))\psi_j
}
These couplings, $g^L$ and $g^R$, involve both mixing  and the
usual $\tau^3-\st Q$ factors. In terms of these,
\bigfirst{\eqn\PhiZfermion{
\Phi_Z=\half(|g^L_{ij}|^2+|g^R_{ij}|^2)
		(I_2(m_i,m_j,M_Z)+ I_3(m_i,m_j,M_Z))
		+ \Re(g^L_{ij}g^{R\ast}_{ji}) I_4(m_i,m_j,M_Z)
}}{\eqnn\PhiZfermion%
$$\eqalign{
\Phi_Z=\half(|g^L_{ij}|^2+|g^R_{ij}|^2)
		(I_2(m_i,m_j,M_Z)&+ I_3(m_i,m_j,M_Z))\cr
		&+ \Re(g^L_{ij}g^{R\ast}_{ji}) I_4(m_i,m_j,M_Z)\cr}
\eqno\PhiZfermion $$}
The expressions for $\Phi_W$ and $\Phi_W^0$ can in fact be read off
from this, but we give them explicitly for future reference. For a
fermion doublet with components of mass $m_1$ and $m_2$,
\eqn\PhiWfermion{
\Phi_W=\quarter(I_2(m_1,m_2,M_W)+I_3(m_1,m_2,M_W))
}
and
\eqnn\PhiWofermion
$$\eqalignno{
\Phi_W^0&=
\lim_{q\to0}{q^2\over4M_W^2}(I_2(m_1,m_2,q)+I_3(m_1,m_2,q)) & \cr
&={1\over2M_W^2(m_1^2-m_2^2)}\left[m_2^4\ln{m_2^2\over\mu^2}
-m_1^4\ln{m_1^2\over\mu^2} +{1\over2}(m_1^4-m_2^4)\right]
& \PhiWofermion \cr}
$$

We have neglected mixing, which is a good approximation in the quark sector.
Were we to neglect mixing in other  sectors, this expression
would be  valid by replacing $1/4\to1/2C({\rm rep})$. In the case of
the wino-zino multiplet, which transforms as the adjoint, the
appropriate replacement is $1/4\to1$. But in addition one has to
include the all important effect of mixing, and the corresponding
formula is just as in eq.~\PhiZfermion.
This involves, in addition,
\eqnn\ifourgoven
$$\eqalignno{
I_4^0&=\lim_{q\to0}(q^2/M_W^2)I_4(m_1,m_2,q)\cr
&=4m_1m_2/M_W^2(m_2^2-m_1^2)[m_2^2\ln m_2^2/\mu^2
 - m_1^2\ln m_1^2/\mu^2- (m_2^2-m_1^2)] & \ifourgoven\cr}$$

It should be kept in mind that the fermions in the neutralino sector
are Majorana. The equations above apply to them too, provided one remembers
to include factors of $1/2$ as appropriate.

Explicit expressions for the integrals appearing in eqs.~\ionedefd\
and~\itwothreedefd\null\ are easily written. We refrain from
doing so because the expressions are lengthy. A detail discussion of
the behavior of these functions can be found in section~8 below.

\newsec{Top and Higgs}
The experimental extraction of gauge couplings in the standard model
depends on assumptions on two other undetermined parameters, the top
and Higgs masses. It is customary to make a global fit of observables
including not just the gauge couplings but these masses ---or at least
the top mass--- as well. How are we to deal with this in our
translation between standard and extended models?

Consider first the dependence on the Higgs mass. In supersymmetric
models one must introduce two distinct Higgs doublets to ensure that
all quarks and charged leptons are massive.  Both fields get vacuum
expectation values, and the physical spectrum is more complicated than
that of the standard model. The details are spelled out in the next
section. For now we would like to concentrate on the question of how
to account for the different Higgs sectors in the standard model and
the supersymmetric one. The problem is that the physical Higgs of the
standard model generally has no simple direct correspondence in the
supersymmetric extension.  The linear combination of fields that gets
a vacuum expectation value, a clear candidate for the corresponding
standard model Higgs, is not a mass eigenstate.

Thus, in computing $\dg_k$ one must be careful to include these
distinctions into $\Delta F_i$ in eq.~\ftildesplit. In fact, $\Delta
F_i$ should include all the supersymmetric Higgs sector contributions
and subtract the standard model ones. However, the standard model
calculation introduces an unknown parameter, the Higgs mass. In fact,
some of the uncertainty in the determination of the standard model's
gauge couplings is due to the lack of determination of the Higgs mass.
In our approach, the standard Higgs contribution would be replaced by
a sum of contributions from the two neutral scalars, plus, in addition,
genuinely new contributions. When the two neutral scalars are
degenerate and with mass equal to that of the standard model's Higgs,
this sum is exactly equal to the standard model Higgs contribution
that it replaces. Because the mass of the Higgs is unknown, and since
the standard model gauge couplings are not very sensitive to it, it is
a good approximation to retain the standard model Higgs contribution,
and instead to neglect the corresponding terms in the model with two
Higgs doublets.

To summarize, given the uncertainties associated with the standard
model determination of gauge couplings, the calculation of Higgs
sector threshold corrections needs include only contributions from
physical scalars in the $W$ and $Z$ self-energy diagrams of \scalardiags.

It must be noted here that in the  run-and-match analysis this
is accomplished by neglecting mixing between the two Higgs
doublets. Then  a correspondence is made between one of them and the
standard model's, while the other gives a new light threshold
correction.

The contribution from the top quark is identical in  the standard
model and its extension. It is included in the function $F_i(\tilde
g_k)$ of eq.~\ftildesplit. The problem is that the determination of
gauge couplings usually uses this as a fit variable. This could lead
us into a complicated analysis of correlations. We are fortunate in
that the models we are interested in ---supergravity unification with
radiative-induced electroweak symmetry breaking--- yield top quark
masses in the range obtained from these fits. Therefore, to good
approximation, the effect of top need not be included in our
calculations.

\newsec{Review of the minimal $SU(5)$ supergravity model.}

In the next section we will illustrate our treatment of the light
thresholds. In preparation we review here the minimal $SU(5)$
supergravity model. We also describe the procedure we have adopted for
the calculation of the spectrum. The particle content of the model
consist of three generation of quarks and leptons, the $SU(3)\times
SU(2)\times U(1)_Y$ gauge bosons, two Higgs doublets and the
supersymmetric particles of all these particles. We neglect the Yukawa
couplings of the two light generations. The superpotential of the
effective low energy theory thus has the form %
\eqn\superpotential{
W=\lambda_tU^cQh_2+\lambda_bD^cQh_1+\lambda_\tau
E^cLh_1+\mu h_1h_2
}
where $Q$ and $L$ are quark and lepton doublet
superfields, and $U^c$, $D^c$ and $E^c$ are the corresponding $SU(2)$
singlets; $h_1$ and $h_2$  are the two Higgs doublets. The soft
supersymmetry breaking terms are  %
%
\bigfirst{
\eqn\vsoft{\eqalign{
V_{\hbox{soft}}=
(&\lambda_tA_tQU^ch_2+\lambda_bA_bQD^ch_1+\lambda_\tau
A_\tau LE^ch_1+B\mu h_1h_2+{\rm h.c.})+\cr
&m_{h_1}^2\vert{h_1}\vert^2+
m_{h_2}^2\vert{h_2}\vert^2+
m_L^2\vert{L}\vert^2+m_{E^c}^2\vert{E^c}\vert^2+
m_Q^2\vert{Q}\vert^2+m_{U^c}^2\vert{U^c}\vert^2+
m_{D^c}^2\vert{D^c}\vert^2.\cr}
}}{
\eqn\vsoft{\eqalign{
V_{\hbox{soft}}=
(&\lambda_tA_tQU^ch_2+\lambda_bA_bQD^ch_1+\lambda_\tau
A_\tau LE^ch_1+B\mu h_1h_2+{\rm h.c.})+\cr
&m_{h_1}^2\vert{h_1}\vert^2+
m_{h_2}^2\vert{h_2}\vert^2+
m_L^2\vert{L}\vert^2+m_{E^c}^2\vert{E^c}\vert^2+\cr
&\qquad\qquad\qquad\qquad
 m_Q^2\vert{Q}\vert^2+m_{U^c}^2\vert{U^c}\vert^2+
m_{D^c}^2\vert{D^c}\vert^2.\cr}
}}

The $SU(5)$ model is specified by the boundary conditions at the
unification scale. The low energy parameters can then be determined
from knowledge of their scale dependence. This is dictated by the
renormalization group.  The renormalization group equations (RGEs) for the
minimal supersymmetric standard model are well known and are given in
appendix~A for completeness.

The boundary conditions at the unification scale that we adopt equate
all of the trilinear coefficients $A_X$ and equate the hard scalar
masses in the soft supersymmetry breaking potential of eqn.~\vsoft.
In addition one has conditions that follow from minimal--$SU(5)$:
equate the tau and beauty Yukawa couplings, $\lambda_b(\MG)
=\lambda_\tau(\MG) $ and equate the gaugino masses $\MM_i$ ($i=1,2,3$,
corresponding to the $U(1)\times SU(2)\times SU(3)$ gauginos). The
free parameters are then the common trilinear coefficients $A$, the
soft Higgs mass parameter $B$, the top Yukawa $\lambda_t$, the
bilinear Higgs mixing coefficient $\mu$, the common gaugino mass
$m_{1\over2}$, the common scalar mass $m_0$, the scale of unification
$\MG$ and the unified gauge coupling $\ag$.  The free parameter
$\lambda_\tau$ is of course fixed by the tau mass.

Given values of these free parameters one can compute the low energy
spectrum of the theory, and then complete the calculation of the
relation between gauge couplings at low energies as outlined in
previous sections. We would like to explore the span of predictions
that correspond to a large region of this free parameter space. It is
quite unnecessary, though, to vary freely over values of all of these
parameters. In computing the spectrum, we take as initial values
$\sin^2\theta_W=0.233$ and $\alpha_s=0.120$, from which we determine
$\MG$ and $\ag$.

The Yukawa couplings of the heavy generation are given by
\eqn\yukawas{
\lambda_{b,\tau}={1\over{\cos\beta}}{m_{b,\tau}\over{v}}
{\hskip 1cm}\lambda_t={1\over{\sin\beta}}{m_{t}\over{v}} }
where $v=\sqrt{v_1^2+v_2^2}=246/\sqrt{2}$~GeV, $\tb=v_2/v_1$ and
$v_{1,2}$ refer to the expectation values of the two Higgs doublets.
We obtain the heavy generation Yukawa couplings by eq.~\yukawas\ from
the input parameters $m_t$ and $\tb$, and the input values of
$m_\tau=1.78$~GeV and $m_b=4.0$~GeV at the $Z$--scale. In our
numerical analysis, we evolve
the Yukawa couplings from the $Z$--scale to $\MG$, and constrain the
parameter space by the requirement $\lambda_b(\MG)$ differs from
$\lambda_\tau(\MG)$ by no more than 5\%. In so doing we neglect the
effect of thresholds on the running of these Yukawa couplings.  Thus
for this procedure we need also a priori knowledge of the angle
$\beta$.

In the next step in our numerical computation of the spectrum we
evolve the Yukawa couplings, the soft SUSY breaking parameters and the
gauge couplings with the boundary conditions specified at $\MG$, to
the electroweak scale, with the RGEs of the MSSM as given in
appendix~A. We then obtain the supersymmetric spectrum at
the weak scale from the equations given below.

In order to span over parameter space at the scale of
grand-unification, and compute the low energy spectrum via the RGE's
subject to the conditions that we obtain the correct masses of the
tau-lepton and bottom-quark and that the correct electroweak breaking
scale is the minimum of the tree level neutral Higgs potential, we
trade the input parameters $|\mu|$ and $B$ for $v$ and $\tb$ (or,
equivalently, $v_1$ and $v_2$). Therefore, $B$ and $|\mu|$ become
computed parameters. This we can do because the running of the SUSY
parameters, as given by the RGE's in appendix~A, does not depend on
the values of $B$ or $\mu$ (except for $\mu$ itself).  Thus, in our
numerical analysis we span over the space generated by $A$, $m_t$,
$m_{1\over2}$, $m_0$, $\tb$ and the sign of $\mu$, and compute the
spectrum for each set of values.

We turn now to a description of the spectrum itself. We lay down here
the basis for our numerical treatment of the next section, making no
attempt to describe the spectrum's salient qualitative features. A
nice account of this can be found in ref.~\martin\ while extensive
numerical analysis of the spectrum can be found in ref.~\groups.
Our results for the sparticle spectrum agree qualitatively with the results
of these references.
For
the two light generation sparticles we neglect the Yukawa couplings in
the RGE's.  The light-generation sparticle masses may then be
analytically calculated from the one-loop RGEs in terms of the three
unknowns $m_{1\over2}$, $m_0$, and $\cos2\beta$:
\eqn\slightmasses{
m^2_{\tilde p}=m^2_{0}+
c_{\tilde p}m^2_{1\over2}+d_{\tilde p}.
}
The coefficients $c_{\tilde p}$ for the different sparticles are
\eqn\cgiven{
c_{\tilde p}=c_3(m_{\tilde p})+c_2(m_{\tilde p})+
Y^2_{\tilde p}c_y(m_{\tilde p})
}
with
\eqna\csgiven
$$\eqalignno{
c_3(m_{\tilde p})=&-{{8}\over9}\left(1-
	\left(1+{{3\ag t}\over{2\pi}}\right)^{-2}\right)&\csgiven a\cr
c_2(m_{\tilde p})=&{3\over2}\left(1-
    \left(1-{{\ag t}\over{2\pi}}\right)^{-2}\right)&\csgiven b\cr
c_y(m_{\tilde p})=&{{10}\over{33}}\left(1-
     \left(1-{{33\ag t}\over{10\pi}}\right)^{-2}\right)&\csgiven c\cr}
$$
where $\ag$ is the coupling at the unification scale and
$t=\log({m_{\tilde p}/{\MG}})$.
The RGE's are integrated to the physical sparticle mass
to obtain the $c$'s.
The $d$'s arise from D--terms in the potential and are given by
\eqn\dsgiven{
d_{\tilde p}=2\left(T_{3_L}^{\tilde p} - {3\over5}Y^{\tilde p}\tan^2\theta_W
\right)\cos2\beta \,M^2_W~. }

In case of small $\tb$ the bottom quark and tau lepton Yukawa couplings
can be neglected as well. In the case of large $\tb$ the heavy
generation sparticle mass matrices are given by
\eqna\sheavies
$$\eqalignno{
M^2_{\tilde t}&=\left(\matrix{
m_Q^2+m_t^2+d_{\tilde u_l}&m_t(A_t+{\mu\;{\cot\beta} })\cr
m_t(A_t+{\mu\;{\cot\beta} })&
m_{U^c}^2+m_t^2+d_{\tilde u_r}\cr}\right) &\sheavies a\cr
M^2_{\tilde b}&=\left(\matrix{
m_Q^2+m_b^2+d_{\tilde d_l}&m_b(A_b+\mu\;{\tb})\cr
m_b(A_b+\mu\;{\tb})&
m_{D^c}^2+m_b^2+d_{\tilde d_r} \cr}\right) &\sheavies b\cr
M^2_{\tilde\tau}&=\left(\matrix{
m_L^2+m_\tau^2+d_{\tilde e_l}&m_\tau(A_\tau+\mu\;{\tb})\cr
m_\tau(A_\tau+\mu\;{\tb})&
m_{E^c}^2+m_\tau^2+ d_{\tilde e_r}\cr}\right) &\sheavies c\cr}$$
The mass eigenstates are obtained by diagonalizing the mass matrices
above by a $2\times2$ unitary transformation.

The chargino and neutralino mass matrices are given by
\eqn\charginoM{
M_{\tilde C}=\left(\matrix{{\tilde
M}_2&M_W\sqrt{2}\sin\beta\cr
			     M_W\sqrt{2}\cos\beta&\mu\cr}\right)~,
}
and
\bigfirst{\eqn\neutralinoM{
M_{\tilde N}=\left(\matrix{
{\tilde M}_1&0&-M_Z\sin\theta_W\cos\beta&M_Z\sin\theta_W\sin\beta \cr
0&{\tilde M}_2&M_Z\cos\theta_W\sin\beta&-M_Z\cos\theta_W\cos\beta \cr
-M_Z\sin\theta_W\cos\beta&M_Z\cos\theta_W\sin\beta&0&\mu \cr
M_Z\sin\theta_W\sin\beta&-M_Z\cos\theta_W\cos\beta&\mu&0 \cr}\right),
}}{\eqn\neutralinoM{
M_{\tilde N}=\left(\matrix{
{\tilde M}_1&0&-M_Z{s_\theta}{c_\beta}&M_Z{s_\theta}{s_\beta} \cr
0&{\tilde M}_2&M_Z{c_\theta}{s_\beta}&-M_Z{c_\theta}{c_\beta} \cr
-M_Z{s_\theta}{c_\beta}&M_Z{c_\theta}{s_\beta}&0&\mu \cr
M_Z{s_\theta}{s_\beta}&-M_Z{c_\theta}{c_\beta}&\mu&0 \cr}\right),
}}
respectively, where ${\MM}_i=(\alpha_i/\alpha_{_U})m_{1\over2}$
$(i=1,2,3)$ are the gaugino masses at their mass
\bigfirst{scale.}{scale, and $s_\theta=\sin\theta_W$,
$c_\theta=\cos\theta_W$, $s_\beta=\sin\beta$ and
$c_\beta=\cos_\beta$.}  The mass eigenstates and mixing can be found
in ref.~\gpw.
%
%

There are two complex Higgs doublets in the MSSM,
$(h_2^+,h_2^0)$, whose neutral
vacuum expectation value (VEV) $v_2$ gives mass to the $2\over3$ charged
quarks, and $(h_1^0,h_1^-)$ whose neutral VEV gives mass to the $-{1\over3}$
charged quarks. Of the eight real degrees of freedom, three are the
Goldstone bosons that are absorbed by the standard model gauge bosons.
The CP--odd neutral Higgs boson $A^0$ and the orthogonal combinations
$h^+=\sin\beta(h_1^-)^*+\cos\beta\;{h_2^+}$,  $h^-=(h^+)^*$
correspond to physical particles with masses
\eqn\cpoddM{
m_A^2=-m_3^2(\tb+\cot\beta)~,
}
and
\eqn\chargedHM{
m_{h^\pm}^2=m_W^2+m_A^2~,
}
where $m_3 = B\mu$.
The two CP--even neutral Higgs bosons mass eigenstates are obtained
by diagonalizing
\eqn\cpevenM{
M_R^2={{\sin2\beta}\over2}\left(\matrix{
\cot\beta{M_Z^2}+\tb{m_A^2} &-(M_Z^2+m_A^2) \cr
-(M_Z^2+m_A^2) &\tb{M_Z^2}+\cot\beta{m_a^2}+\Delta \cr}\right)~,
}
where $\Delta$ is the one loop radiative correction to the neutral Higgs boson
masses\okada
\eqn\deltahiggsM{
\Delta={3\over{8\pi^2}}{{g^2m_t^4\cot\beta}\over{m_W^2}}\log
{\left(1+{m^2_{sq}\over{m_t^2}}\right)}~.
}
The linear combination of Higgs fields that gets a
VEV, $h_{sm}$, and the one that does not get  a VEV, $h_{susy}$, are obtained
from the mass eigenstates $h$ and $H$ by a rotation,
\eqn\higgsrotation{
\left(\matrix{h_{sm}\cr h_{susy}\cr}\right)=
  \left(\matrix{\cos(\alpha-\beta)&\sin(\alpha-\beta)\cr
                -\sin(\alpha-\beta)&\cos(\alpha-\beta)\cr}\right)
  \left(\matrix{h\cr H\cr}\right)~,
}
where $\a$ is the angle that rotates from the gauge to the mass eigenstates.
Requiring a negative eigenvalue for the neutral Higgs mass--squared
matrix and that the Higgs potential is bounded from below imposes two
conditions on the running of the mass parameters:
$$\eqalignno{&1.~~m_1^2m_2^2-m_3^4<0\cr
  &2.~~m_1^2+m_2^2-2\vert{m_3^2}\vert>0\cr}$$
where $m_{1,2}^2=m_{h_{1,2}}^2+\mu^2$ and $m_3^2=B\mu$.

\newsec{Numerical Results and Discussion}
We first consider the functions $\Phi_X$ of section~5 as compared with
the naive (leading-log, or "run-and-match") result. The latter can be
obtained by the standard run-and-match computation, or simply by
taking the large $m/M_Z$ limit of our expressions. Both procedures must
agree when in our expressions $SU(2)\times U(1)$ breaking is neglected.

Consider the contributions of scalars to $\dl$ of
eqs.~\dlighti\ and~\dlightii, both in our procedure and using the run-and-match
approximation. When $SU(2)\times U(1)$ breaking is neglected, so that
all members of a weak-isospin multiplet are degenerate, all three
functions $\Phi_X$ are simply proportional to the function $I_1$.
Thus, for the contribution of any one scalar, the ratio of
$\delta_{\rm light}$ computed in the run-and-match approximation to
our result is simply the ratio of the leading-log term in $I_1$ to the
whole expression for $I_1$. This ratio is plotted in
\fig\ratioscalars{Ratio of the naive run-and-match approximation to
the complete one loop threshold correction functions $\Phi_X$, due to
a virtual scalar.}\  as a function of $m/M_Z$.
It is apparent that the error in the approximation is small except
for $m/M_Z\approx1$.

Next, consider the contributions of fermions to $\dl$.
The $\Phi_g$ piece is easiest to analyze, since it is simply
proportional to the function $I_2$; see eqs.~\itwothreedefd{a}
and~\Phigfermion. In \fig\gratiofermions{Ratio of the naive
run-and-match approximation to the complete one loop threshold
correction function $\Phi_g$, due to virtual fermions.}\ we plot the
ratio of the naive run-and-match approximation to the complete one
loop threshold correction function $\Phi_g$, due to virtual fermions.
In this case, again, the convergence to the naive run-and-match limit
is extremely fast.

The most interesting case is that of the fermion contributions to
$\Phi_Z$ and $\Phi_W$. Recall, from section~6 that $\Phi_Z$ and $\Phi_W$ are
linear combinations of the integrals $I_2$, $I_3$ and $I_4$.
The behavior of $I_2$ was analyzed
above, and was found to be quite uninteresting. Turning our attention
to $I_3$, we see by inspection of eq.~\itwothreedefd{b}\ that its
asymptotic behavior, for large $m$, is
$I_3(m,m,q)\sim-4m^2/q^2\ln(m^2/\mu^2)+2/3+q^2/15m^2+{\cal
O}(q^4/m^4)$. This is an example of {\it non-decoupling}, in which the
leading term grows as a power (rather than a log) of the mass. In
fact, the combination $\Phi_Z-\Phi_W$ is precisely what enters the
computation of Veltman's $\rho$ parameter, which is well known to grow
as $\sim m_t^2$ in the standard model\velt. Notice though that for
particles in vectorial representations the prefactor $(\tau_L^3-
\tau_R^3)^2$ vanishes. So a large effect will depend on the breaking
of $SU(2)\times U(1)$.

It is interesting to note that the combination $I_3+I_4$ which often
enters into $\Phi_Z$, vanishes in the limit of degenerate multiplets.
This can be seen by writing it as
\bigfirst{\eqn\ifour{
I_3(m_1,m_2,q)+I_4(m_1,m_2,q)= -{4\over q^2}(m_2-m_1)\int_0^1 dx \,
(xm_2-(1-x)m_1)\log M^2(x)/\mu^2~.
}}{\eqn\ifour{\eqalign{
I_3(m_1,m_2,q)&+I_4(m_1,m_2,q)=\cr
&-{4\over q^2}(m_2-m_1)\int_0^1 dx \,
(xm_2-(1-x)m_1)\log M^2(x)/\mu^2~.\cr}}}

These type of leading corrections to $\dl$ are missed altogether by
the naive run-and-match calculation.  In fact, there is no pure log
term in $I_3$, an indication that it is invisible to the naive
approach. And, clearly, the combination $I_3+I_4$ is missed altogether
too.  Therefore we cannot quantify the effects of these terms,
relative the run-and-match approach, in a manner similar to the above
examples. We postpone further consideration until we come to
computations in the full $SU(5)$ supergravity model, below.

To get some idea of how important the $SU(2)$ breaking effects can be,
we can look at the contribution to $\dl$ from a doublet of scalars,
with masses $m_1$ and $m_2$ respectively, and compare to the would be
contribution from a degenerate doublet with some average mass, $\ma$.
For definiteness we shall take $\ma=(m_1+m_2)/2$, although one could
just as well use, say, $\ma=\sqrt{m_1m_2}$. In \fig\sutwobreak{The
effect of $SU(2)$ breaking, as included in our analysis, in the scalar
contribution to $\Phi_W$. This is a plot of the ratio
$I_1(m_1,m_2,q)/I_1(\ma,\ma,q)$, where $\ma=(m_1+m_2)/2$, for
$m_2/q=1.5,3.5,5.5$ and~$7.5$ (corresponding to the ever increasing
location of the peaks in the graph).}\ we plot the ratio
$I_1(m_1,m_2,q)/I_1(\ma,\ma,q)$ for $m_2/q=1.5,3.5,5.5$ and~$7.5$
(corresponding to the ever increasing location of the peaks in the
graph). It is clear that $SU(2)$ breaking effect can be rather
considerable. It should be pointed out that large $SU(2)$ breaking
effects would show up as significant deviations from unity of
Veltman's $\rho$-parameter. For consistency with the observed value
of $\rho$, the parameters of the SUSY-GUT are such that the $SU(2)$
breaking effects are small.

These comparisons, although instructive, are incomplete. They miss the
possibility of  cumulative effects of the many particles in the rich
supersymmetric spectrum.  We therefore use our equations in the
context of the model described in the previous section. This will
allow consideration of the effects of mixing. These are obviously
important and altogether missed by the run-and-match approach, but
serious consideration of them requires a choice of reasonable mixing.
In the minimal supergravity model based on $SU(5)$ the mixing is
completely fixed in terms of other more fundamental parameters.

We calculate the spectrum in the minimal supersymmetric $SU(5)$ model
as described in section~7. There are five input parameters, namely
$A$, $m_0$, $m_{1\over2}$, $\tb$ and the top quark mass $m_t$. In
addition one must specify the sign of $\mu$. They
are constrained by the requirement of radiatively induced symmetry
breaking of the electroweak group, and lower bounds on supersymmetric
particle masses. We perform a systematic, although rather
coarse-grained, search through parameter space. Each of the five
parameters is allowed to vary over a reasonable range. The range is
sampled at some fixed predetermined interval size. Table 1 summarizes
the range and size of the sampling interval for each of the five
parameters. This represents a sampling of more than $4\times10^4$
points.
Of the resulting spectrum we require that it
does not violate experimental bounds: we require that chargino,
squark, gluino, slepton and neutralino masses to be in excess of 45
GeV, 100 GeV, 150 GeV, 43 GeV and 20 GeV, respectively. We also require
that the lightest neutralino is lighter than the chargino.
Only some
$3\times10^4$ sampling points in parameter space survive this requirements.
To study the dependence on each of the input parameters we use a finer grid
for one of the inputs and take two or three extreme points for the
other input parameters.
We find that $m_t<160$~GeV and $\tan\beta<30$. This
is the case in part because we have assumed that the tau and bottom
Yukawa couplings are equal at the unification scale. The gluino mass limit,
$M_{\tilde g}>150$~GeV, constrains $m_{1\over2}$ to be above 55 GeV.
Upper limits on $\vert{A_0}\vert$,
$m_0$ and $m_{1\over2}$ can be set by resorting
to naturalness arguments~\ross. However, as our interest is in the SUSY
spectrum close to the electroweak scale, we impose upper
bounds on these parameters not far above the electroweak scale.
\vskip2cm

{\hbox
{\hfill
{\begintable
\  \ \|\ 	    $X_i$ \ \|\ $X_f$ \ \|\ $\Delta X$ \crthick
$A_0$ 	 	\|\ -200  \ \|\	200   \ \|\    50  \nr
$m_0$ 	 	\|\ ~~~0  \ \|\ 400   \ \|\    50  \nr
$m_{1\over2}$ 	\|\ ~100  \ \|\	300   \ \|\    25  \nr
$\tb$ 		\|\ ~~~2  \ \|\	~30   \ \|\    ~4  \nr
$m_t$ 		\|\ ~110  \ \|\	170   \ \|\    10  \endtable}
\hfill}
\bigskip
\parindent=0pt
{\it Table 1. The range and sampling size of the parameter space.
Each free parameter $X$ is sampled in the interval $(X_i,X_f)$ with
spacing $\Delta X$ between consecutive points.}
}\vskip 2cm

How good is the run-and-match analysis in this case? In
\fig\sinvssin{Plot of the calculated $\st$, this work, in (a) method--I, and
(b) method--II, versus the calculated $\st$ in the standard
`run-and-match' approximation\ekn. See eq.~\dlightdefd. Each point on
this plot represents a specific choice of free parameters of the
minimal supersymmetric $SU(5)$ grand-unified theory that survives
experimental bounds on the spectrum. The parameter space spanned here
is detailed in Table~1.}  we compare our calculations, in (a)
method--I, and (b) method--II, with the result of ref.~\ekn. We plot
$\st$, calculated in~\dlightdefd\ and~\dlighti\ or~\dlightii\ using
$\as=0.120$ and $\ae=1/127.9$, against $\st$ in the run-and-match
approximation.  Each point on this plot represents a specific choice
of the free parameters. It is clear that for many points the
run-and-match approximation is far too crude to be useful. The effect
of two-loop running can be incorporated in these graphs (and the
following ones) by a common shift of all points. The two band
structure that can be seen in \sinvssin\ come from the two signs of
$\mu$, the lower higher values corresponding to positive $\mu$. One
should not worry much about the discrepancy between the two graphs in
\sinvssin\  since they correspond to different interpretations of
the same ``experiment''.

\nfig\sinvsmC{Scatter plot of $\st$ against the mass of the lightest
chargino, $M_{C,{\rm min}}$, for the minimal supersymmetric
$SU(5)$ grand-unified theory, in (a) method--I, and
(b) method--II. Each point represents a choice of
parameters; see Table~1.}
\nfig\sinvsmH{Scatter plot of $\st$ against the mass of the lightest
neutral Higgs, $M_{H,{\rm min}}$, for the minimal supersymmetric $SU(5)$
grand-unified theory, in (a) method--I, and
(b) method--II. Each point represents a choice of parameters;
see Table~1.}
\nfig\sinvsmT{Scatter plot of $\st$ against the mass of the lightest
top-squark, $M_{T,{\rm min}}$, for the minimal supersymmetric $SU(5)$
grand-unified theory, in (a) method--I, and
(b) method--II. Each point represents a choice of parameters;
see Table~1.}
\nfig\sinvsmN{Scatter plot of $\st$ against the mass of the lightest
neutralino, $M_{N,{\rm min}}$, for the minimal supersymmetric $SU(5)$
grand-unified theory, in (a) method--I, and
(b) method--II. Each point represents a choice of parameters;
see Table~1.}
\nfig\sinvsrho{Scatter plot of $\st$ against the deviation of
Veltman's $\rho$ parameter from unity, for the minimal supersymmetric
$SU(5)$ grand-unified theory in method--I. Each point represents a
choice of parameters; see Table~1.}
In \figs{\sinvsmC - \sinvsmN}\ we plot the dependence of $\st$ in our
calculation on the mass of a few of the superparticles, in (a)
method--I, and (b) method--II. The dependence on the lightest chargino
mass, $M_{C,{\rm min}}$, can be seen in \sinvsmC. As expected, the
points tend to concentrate around the $\st$ value of the run-and-match
approximation as $M_{C,{\rm min}}$ increases. The mass of the lightest
neutral Higgs increases rapidly as a function of $\tb$, reaching a
limiting value of about $100$~GeV at $\tb\sim6$. This, rather than the
two signs of $\mu$,  is the reason
for the two band structure of \sinvsmH, a scatter plot of $\st$
against $M_{H,{\rm min}}$. The band at about $60$~GeV is made up
mostly of points at $\tb=2$, while the band at $100$~GeV consists
mostly of points with $\tb\ge6$.  Fig.~11 
shows $\st$ against the mass of the lighter of
the two top-squarks. It is similar to \sinvsmC\ in that points tend to
accumulate at the run-and-match value as $M_{T,{\rm min}}$ increases,
although the spread seems larger.  Fig.~12 
again demonstrates these effects, now as a function of the lightest
neutralino. Finally, in \sinvsrho\ we verify that $\delta\rho$ remains
small in method--I. Note that the larger values of $\delta\rho$
correspond to a narrow band of $\st$.

We would like to emphasize that our numerical analysis is not intended
as a complete analysis of the parameter space and of the SUSY
spectrum. Nor do we believe, in view of the huge parameter space, that
such analysis carry substantial meaning.  The parameter space can be
constrained by making specific assumptions, motivated to some extent
by theoretical considerations, on the form of the K\"ahler potential
in the supergravity model, and consequently on the initial boundary
conditions at the unification scale.  Partial analysis of the SUSY
spectrum, based on such specific assumptions, have been performed by
several groups\groups.  In this paper, our purpose is to address the
question of how to appropriately account for the effects of light
thresholds, in a way that can consistently incorporate two-loop
running, and to illustrate the need for a proper method to do so.  Our
numerical analysis should be regarded merely as illustrative.

\newsec{Conclusions}
The standard run-and-match approach to the calculation of light
threshold effects is seen to be inaccurate when the supersymmetric
spectrum is not too heavy and there is substantial $SU(2)$ breaking
and mixing. We have described a better approximation which
incorporates these effects. Moreover, our method incorporates readily
the effects of two-loop running. Although we have concentrated on the
predicted value of $\st$ in the numerical analysis in this work,
clearly similar analysis apply to the calculation of other quantities
of interest. For example, one may include threshold effects in the
calculation of the spectrum itself \hrs.

The calculation that we have presented is a sensible approach,  as
long as supersymmetric particles are not found and provided there is
no evidence of experimental deviations from standard model
expectations.  When either of these assumptions is violated the
calculation should be replaced by a full calculation of the novel
process, including possible non-universal effects, that is,
non-oblique corrections.

In the mean time one may use this type of calculation to restrain more
severely the allowed parameter space of supersymmetric grand-unified
theories. It would be interesting to produce an analysis including
simultaneously the predictions of $\st$ and other restricting
quantities, to better limit the boundaries of parameters space. For
example, the inclusive rate for radiative $B$ decays may be fairly
restrictive of extreme values of $\tb$. Unfortunately, to the
extent that the standard model is successful, one can always find a
region of the SUSY-GUT that satisfies all experimental constraints, for
the sparticles decouple as they become much heavier than the massive
vector bosons, and in that limit the predicted value of $\st$ is
consistent with experiment.

\vskip1in
\newsec{Acknowledgments}
We would like to thank R. Arnowitt for pointing out an error in the
calculation in section~5 in an earlier version of the manuscript, and
H. Anlauf for pointing out an error in the numerical code and for his
help in improving it. Thanks to W. Bardeen for extensive
discusions. One of us (BG) would like to thank the Alfred P. Sloan
Foundation for partial support. AF would like to thank the Feinberg
school for partial support and the SSC laboratory for its hospitality
while part of this work was conducted.  This work is supported in part
by the Department of Energy under contract DE--AC35--89ER40486.

\appendix{A}{ RGE's for the MSSM }
The renormalization group equation for the minimal supersymmetric
standard model are well known and are given here for completeness.

%
%
\eqna\RGE
$$\eqalignno{
{{d\lambda_t}\over{dt}}&={\lambda_t\over{\pie1}}
      (3\lambda_t^2+{1\over2}\lambda_b^2-
                {8\over3}g_3^2-{3\over2}g^2_2-{{13}\over{30}}g^2_1)&\RGE a\cr
{{d\lambda_b}\over{dt}}&={\lambda_b\over{8\pi^2}}
      (3\lambda_b^2+{1\over2}\lambda_t^2+{1\over2}\lambda_\tau^2-
                {8\over3}g_3^2-{3\over2}g^2_2-{{7}\over{30}}g_1^2),&\RGE b\cr
{{d\lambda_\tau}\over{dt}}&={\lambda_\tau\over{8\pi^2}}
      (2\lambda_\tau^2+{3\over2}\lambda_b^2-
                {3\over2}g^2_2-{{9}\over{10}}g_1^2),&\RGE c\cr
{{d\mu}\over{dt}}&={\mu\over{8\pi^2}}
      ({3\over2}\lambda_t^2+{3\over2}\lambda_b^2+{1\over2}\lambda_\tau^2-
                {3\over2}g^2_2-{{9}\over{10}}g_1^2),&\RGE d\cr
{{dA_t}\over{dt}}&={1\over{4\pi^2}}
      (3\lambda_t^2A_t+{1\over2}\lambda_b^2A_b+
   	{8\over3}g^2_3\MM_3+{3\over2}g^2_2\MM_2+
                           {{13}\over{30}}g_1^2\MM_1),&\RGE e\cr
\bigfirst{
{{dA_b}\over{dt}}&={1\over{4\pi^2}}
      (3\lambda_b^2A_b+{1\over2}\lambda_t^2A_t+{1\over2}\lambda_\tau^2A_\tau+
 {8\over3}g^2_3\MM_3+{3\over2}g^2_2\MM_2+{{7}\over{30}}g_1^2\MM_1),&\RGE f\cr}%
{     {{dA_b}\over{dt}}&={1\over{4\pi^2}}
      (3\lambda_b^2A_b+{1\over2}\lambda_t^2A_t+{1\over2}\lambda_\tau^2A_\tau+
  {8\over3}g^2_3\MM_3+{3\over2}g^2_2\MM_2+{{7}\over{30}}g_1^2\MM_1),\cr
 & &\RGE f\cr}
{{dA_\tau}\over{dt}}&={1\over{4\pi^2}}
      (2\lambda_\tau^2A_\tau+{3\over2}\lambda_b^2A_b+
   	{3\over2}g^2_2\MM_2+{{9}\over{10}}g_1^2\MM_1),&\RGE g\cr
{{dB}\over{dt}}&={1\over{4\pi^2}}({3\over2}\lambda_t^2A_t+
                 {3\over2}\lambda_b^2A_b+{1\over2}\lambda_\tau^2A_\tau+
   {3\over2}g^2_2\MM_2+{{3}\over{10}}g_1^2\MM_1),&\RGE h\cr
{{dm^2_{h_1}}\over{dt}}&={1\over{8\pi^2}}(3F_b+F_\tau-
   3g^2_2\MM_2^2-{{3}\over{5}}g_1^2\MM_1^2),&\RGE i\cr
{{dm^2_{h_2}}\over{dt}}&={1\over{8\pi^2}}(3F_t-
   3g^2_2\MM_2^2-{{3}\over{5}}g_1^2\MM_1^2),&\RGE j\cr
{{dm_Q^2}\over{dt}}&={1\over{8\pi^2}}(F_b+F_t-
   {{16}\over3}g^2_3\MM_3^2-
           3g^2_2\MM_2^2-{{1}\over{15}}g_1^2\MM_1^2),&\RGE k\cr
{{dm_{U^c}^2}\over{dt}}&={1\over{8\pi^2}}(2F_t-
   {{16}\over3}g^2_3\MM_3^2-{{16}\over{15}}g_1^2\MM_1^2),&\RGE l\cr
{{dm_{D^c}^2}\over{dt}}&={1\over{8\pi^2}}(2F_b-
   {{16}\over3}g^2_3\MM_3^2-{{4}\over{15}}g_1^2\MM_1^2),&\RGE m\cr
{{dm_{L}^2}\over{dt}}&={1\over{8\pi^2}}(F_\tau-
   3g^2_2\MM_2^2-{{3}\over{5}}g_1^2\MM_1^2),&\RGE n\cr
{{dm_{E^c}^2}\over{dt}}&={1\over{8\pi^2}}(2F_\tau-
   {{12}\over{5}}g_1^2\MM_1^2),&\RGE o\cr
}$$
where,
\eqna\Fts
$$\eqalignno{
F_t&=\lambda_t^2(m_Q^2+m_{U^c}^2+m_{h_2}^2+A_t^2)&\Fts a\cr
     F_b&=\lambda_b^2(m_Q^2+m_{U^c}^2+m_{h_1}^2+A_b^2)&\Fts b\cr
  F_\tau&=\lambda_\tau^2(m_L^2+m_{E^c}^2+m_{h_1}^2+A_\tau^2)&\Fts c\cr
}$$

 \listrefs \listfigs
\bye